\documentclass[journal=jacsat,manuscript=article]{achemso}
\setkeys{acs}{articletitle=true}
\setkeys{acs}{maxauthors=10}
\setkeys{acs}{etalmode=truncate}

\usepackage{amssymb}
\usepackage{amsmath}
\usepackage{graphicx}
\usepackage{mathtools}
\usepackage{bm}
\usepackage{color}
\usepackage{longtable}
\usepackage{chemfig}
\usepackage[version=4]{mhchem}

\SectionNumbersOn

\title{Finding the Right Bricks for Molecular Lego:\\
A Data Mining Approach to Organic Semiconductor Design}

\author{Christian Kunkel}
\author{Christoph Schober}
\author{Johannes T. Margraf}
\author{Karsten Reuter}
\author{Harald Oberhofer}
\email{harald.oberhofer@tum.de}
\affiliation{Chair for Theoretical Chemistry and Catalysis Research Center, Technische Universit\"at M\"unchen, Lichtenbergstra\ss{}e 4, D-85747 Garching, Germany}

\begin{document}

\begin{abstract}
Improving charge carrier mobilities in organic semiconductors is a challenging task that has hitherto primarily been tackled by empirical structural tuning of promising core compounds. Knowledge-based methods can greatly accelerate such local exploration, while a systematic analysis of large chemical databases can point towards promising design strategies. Here, we demonstrate such data mining by clustering an  in-house database of $>64.000$ organic molecular crystals for which two charge-transport descriptors, the electronic coupling and the reorganization energy, have been calculated from first principles. The clustering is performed according to the Bemis-Murcko scaffolds of the constituting molecules and according to the sidegroups with which these molecular backbones are functionalized. In both cases, we obtain statistically significant structure-property relationships with certain scaffolds (sidegroups) consistently leading to favorable charge-transport properties. Functionalizing promising scaffolds with favorable sidegroups results in engineered molecular crystals for which we indeed compute improved charge-transport properties.
\end{abstract}

%This heralds the value of data-based approaches as basis for a future targeted design of materials for organic electronics.

\maketitle

\section{Introduction}

In the last few years, organic semiconductors have received considerable attention due to their comparatively small ecological and economical footprint, as well as due to their great versatility promising a wide spectrum of novel material properties.\cite{Minemawari2011NAT,Stavrinidou2015SCA,Xu2017SCI,Nikolka2017NM,Wang2018CSR}  Their potential uses include organic field effect transistors (OFETs)\cite{Wang2012CR}, photovoltaics (OPVs),\cite{Lin2012CSR} light emitting diodes (OLEDs)\cite{Xu2016JMCC} or even sensors\cite{Anikeeva2015SCI}. A major limitation to the commercial application of this promising class of materials has been the failure to reproducibly yield high charge-carrier mobilities as a prerequisite for the electronic performance.\cite{Anthony2014NM,Yavuz2017PCCP} Thanks to intense research efforts,\cite{Venkateshvaran2014Nature,Ostroverkhova2016CR,Kang2017NM,Bredas2017NM} a number of materials now exist with mobilities exceeding even that of amorphous silicon.\cite{Kim2014JACS} Yet, the sheer endless materials spaces constituted by assembling building blocks of functionalized organic molecules raise the suspicion that these few materials are nothing but the top of an iceberg.

So far, experimental discovery of new materials commonly follows a cycle of synthesis, device manufacture, performance evaluation and subsequent molecular tuning, based on chemical intuition, experience and---to a certain degree---trial and error.\cite{Wang2012CR,Mei2013JACS,Wang2018CSR} Theoretical investigation is then often used to understand and exploit the key charge-transport mechanisms \cite{Oberhofer2012PCCP,Zhang2017ACR,Sosorev2017PCCP, Reig2017PCCP,Li2018JACS}. Although highly successful, such efforts are limited by the pace of iterative improvement, while only locally exploring the chemical space around a compound family. As a more recent development, a number of groups have therefore started to conduct {\em in silico} materials discovery calculations,\cite{Olivares2011EES,Akimov2015CR,Schober2016JPCL, Mercado2018CM, Mayer2018ChemSci, Meredig2014PhysRevB} to guide synthetic studies to promising targets. In these screening approaches, readily available or easily calculable properties of the system---so called descriptors---which are correlated with critical observables of the system (such as the carrier mobility) are determined for a larger set of candidate materials. These then yield a ranking of the screened candidate materials based on their suitability as an organic semiconductor.

In other fields of science, such as heterogenous catalysis,\cite{Norskov2009NC,Andersen2017ACSCat, Goldsmith2018AlChe} superhard materials,\cite{Mansouri2018JACS} or drug discovery,\cite{CHEN2018DrugDiscToday}, such screening studies are already followed up with modern large-scale data mining or even machine learning approaches to extract more general design criteria.\cite{Zunger2018NatRevChem} Corresponding approaches undertaken for organic semiconductors (OSCs) have so far been limited to a much smaller scale.\cite{Lillienfeld2011JCTC, Sahu2018AdvEnMater} In this work, we now take this step and combine the statistical tools of modern data mining approaches with the molecular analysis methods known as cheminformatics to uncover guiding principles for the design of organic semiconducting materials.

\subsection{Descriptor-Based Data Mining of OSCs}

Two of the most prominent charge-transport descriptors for organic semiconductors are the network of electronic couplings $|H_\text{ab}|$ between microscopic sites involved in the charge transport and the reorganization energy $\lambda$, which measures the cost of accommodating a new charge state after the carrier has moved to the next site. While $|H_\text{ab}|$ and $\lambda$ are most closely associated with the small polaron hopping mechanism of carrier transport known from Marcus theory,\cite{Marcus1956JCP,Marcus1993RMP} both also significantly influence the carrier mobility in other transport regimes.\cite{Oberhofer2017CR} In evaluating $\lambda$, usually only the dominant local, intra-molecular reorganization energy is considered.\cite{McMahon2010JPCL} In contrast to $|H_\text{ab}|$, the reorganization energy can then be estimated cheaply as a single site property, while at the same time decisively---in Marcus-like models even exponentially---determining the carrier mobility.\cite{Hutchison2005JACS,Oberhofer2012PCCP}

A number of screening studies thus initially focused on the optimization of $\lambda$ through modification of molecular sidegroups in, for example, dinaphtho-thienothiophene (DNTT),\cite{Sokolov2011NC} triarylamine-based donor molecules for solar cells,\cite{Alberga2017CM} or high-performance polymer semiconductors.\cite{Li2012SR} Taken together, the studies impressively demonstrated how {\em in silico} exploration of material spaces can efficiently identify high-performance organic semiconductors. While similar approaches are now also starting to surface in experimental studies,\cite{Liu2018ACR} most recent {\em in silico} design studies start to optimize $\lambda$ in newly generated derivates. The design problem is thereby mainly tackled by construction and computational screening of highly focused molecular libraries based on a number of different strategies. These range from the generation of polymer semiconductors from preselected building blocks\cite{Moral2017JPCC, Hutchison2005JACS} to the optimization of $\lambda$ through heteroatom replacement\cite{ZHU201616, oshi-anthracene-tetracene-substitution}, introduction of sidegroups\cite{Hutchison2005JACS, oshi-naphthalene-substitution, oshi-anthracene-tetracene-substitution, nonbonding_lambda_1, substitution_effects_bredas, Lee2010, VEDOVA-BROOK2004-Chem_Phys, ZHAO201325, PAN2006CompMaterSci} or ring fusion\cite{nonbonding_lambda_2, Ming2009}.

Modern data-driven approaches instead pursue a complementary strategy. Rather than screening a focused library for potential candidate materials, in particular data mining techniques unveil correlations in large data sets and thereby provide leads for a rational optimization beyond the originally considered design space\cite{Bing2018ACSNano, Gomez-Bombarelli2016NatMater, Agrawal2016APLMater}. Size and especially diversity of the underlying data source are thus key for the generality and transferability of the identified correlations. A suitable representation of the data critically determines the efficiency with which these correlations can be extracted, whether by machine learning or more traditional fitting techniques. In several fields, development and application of such knowledge-based approaches has already led to a greatly accelerated materials design\cite{Butler2018Nature, collection_ml_data_science_in_materials, Tabor2018NatMaterRev}.

Here, we show, that a  data-mining rooted in the chemical understanding of the structure---often termed chemical intuition---can indeed yield general design principles for organic semiconductors. To this end, we concentrate on the class of crystalline molecular organics and analyze an in-house dataset containing $> 64.000$ experimentally characterized mono-molecular crystals. This dataset was originally assembled to computationally screen for hitherto unknown organic crystals with favorable charge-transport properties.\cite{Schober2016JPCL} As further described below, it contains an unprecedented molecular diversity and a wealth of promising semiconductors as identified on the basis of computed high electronic couplings $|H_{ab}|$ and low reorganization energies $\lambda$. In this work, we now perform a two-step data mining procedure on this dataset. Using the concept of Bemis-Murcko scaffolds\cite{BM-Paper} that has already been successfully used in pharmaceutical research, we establish a useful data representation that uncovers statistically significant performance differences between the individual molecular building blocks of the crystals. Clustering the data first according to their Bemis-Murcko scaffolds, i.e. the molecular backbones, we find clear structure-property relationships, with a range of scaffolds consistently leading to favorable charge-transport properties, despite the positional disorder brought about by differing sidegroups and anchor points on them. In a second step, we then cluster the data according to the sidegroups that are attached to the molecular scaffolds. Again, we obtain clear relationships with a range of sidegroups generally lowering the reorganization energy, regardless of the scaffold they are attached to. These findings suggest the combination of certain, high-coupling scaffolds with reorganization energy-reducing sidegroups as a promising design criterion for optimized charge transport in organic molecular crystals. We validate the transferability and accuracy of this criterion by a corresponding sidegroup-engineering of molecular testsets and indeed find significant improvements in the charge-transport properties.

\section{Methods}

\subsection{Dataset}
Our study is based on the ``64k-dataset''\cite{Schober2016JPCL}, containing information on 64.725 organic crystals composed of 61.770 unique molecular structures, extracted from the Cambridge Structural Database\cite{CSD} and screened for favorable charge-transport related descriptors $\lambda$, $|H_{ab}|$. All crystals are mono-molecular, i.e. they contain only a single type of molecule. For a cheminformatics based analysis, $xyz$-coordinates of the molecular structures were extracted from the crystal and converted to canonical smiles strings \cite{Weininger1989} using Openbabel\cite{Openbabel}. All subsequent molecular analysis was carried out using the RDKit.\cite{rdkit} Marvin was used for drawing chemical structures\cite{Marvin}. Pymatgen was used to analyze the occurrence of molecular point groups and crystal space groups. \cite{ONG2013PymatgenCompMaterSci}

\subsection{Electronic Structure Calculations}
The computational settings and methodology were described in detail in the preceding work that established the ``64k-dataset''. \cite{Schober2016JPCL} Thus, we here only briefly summarize them for completeness. Charge-transport related descriptors $\lambda$ and $|H_{ab}|$ were computed using first-principles Density-Functional Theory (DFT) at the GGA (BLYP\cite{Becke,LYP}) level of theory. All calculations were done using the FHI-aims package \cite{BLUM20092175,AIMS2} employing its well-established basis-sets and integration grid settings, detailed below in this section.
Our work on the ``64k-dataset'' showed that even semi-local DFT was able to recover trends among different crystals for both $|H_{ab}|$\cite{Schober2016JCP} and $\lambda$.\cite{Schober2016JPCL} For each crystal in the dataset, the nearest-neighbor molecular dimers were first extracted by a purpose-built python program. Electronic couplings $|H_{ab}|$ for each dimer were then calculated using the fragment molecular orbital approach (FO-DFT)\cite{FODFT-original} in the $H^{2n-1}@D+A$ variant.\cite{Schober2016JCP} The electronic wave functions were expanded in an extended ''tier 1'' basis set of FHI-aims using light integration settings. The highest electronic coupling among all dimers was extracted from this data and constitutes the $|H_{ab}|$ entering the dataset.\cite{Schober2016JPCL} Keeping in mind that the ``64k-dataset'' consists solely of crystals composed of a single species of molecule each, the internal reorganization energy was evaluated for each molecule based on the 4-point scheme,\cite{Nelsen1987JACS} using a non-periodic QM/MM embedding setup.\cite{Schober2016JPCL} In this ONIOM-based scheme,\cite{Svensson1996JPC} a shell of nearest neighboring molecules mimics the steric hindrance due to a realistic solid-state environment during geometry optimization of a central molecule. The shell of nearest neighbours was thereby extracted based on the minimal distance found between the atoms of any two molecules in the crystal. All molecules that approach the central molecule to within this distance times $1.5$ were included in the neighbor shell. Then, keeping the shell of nearest neighbors fixed, the local structure optimisation of the central molecule was performed using the Broyden-Fletcher-Goldfarb-Shanno (BFGS) implemented in the Atomic Simulation Environment\cite{ASE2017} using a convergence criterion $\rm{f_{max} <}$ 0.05 eV/\AA. The MM parts of these calculations were based on the LAMMPS package\cite{Plimpton1995jcph} with the ''Universal Force Field'' (UFF)\cite{Rappe1992JACS} and the QEq charge equilibration scheme\cite{Rappe1991JPC} in order to determine interaction parameters and fixed partial charges, respectively. To ensure maximum accuracy of our main descriptor, we re-calculated all reorganization energies $\lambda$ in the present database using default ''tight'' settings and the respective ''tier 2'' basis-sets of FHI-aims.

\subsection{Statistical Methods}
We used the two-tailed Mann-Whitney-U test\cite{Mann1947} to assess whether two samples are drawn from the same distribution (null hypothesis) or not (alternative hypothesis). Note that the Mann-Whitney-U test is non-parametric and therefore avoids assumptions on the underlying distributions. To account for the great number of statistical tests we carried out, we corrected the computed $p$-values for each descriptor using the false discovery rate correction of Benjamini-Hochberg,\cite{Benjamini1995} as implemented in the statsmodels package\cite{Skipper-Statsmodels}. These corrected $p$-values are then named $q$-values. Results were marked statistically significant, if the $q$-value was found to be below 0.05, i.e.~if the probability that the statistical testing wrongly rejects the null hypothesis is smaller than 5\,\%. Similar methods and criteria are commonly used in Gene Set Enrichment Analysis\cite{gene-set-enrichment-analysis} and have also been applied to cheminformatics\cite{compound-set-enrichment, drug-set-enrichment-analysis} for the identification of active compound series.

\section{Results}

\subsection{Molecular Scaffolds}

\begin{figure}[ht]
 \centering
  \includegraphics[width=.7\textwidth]{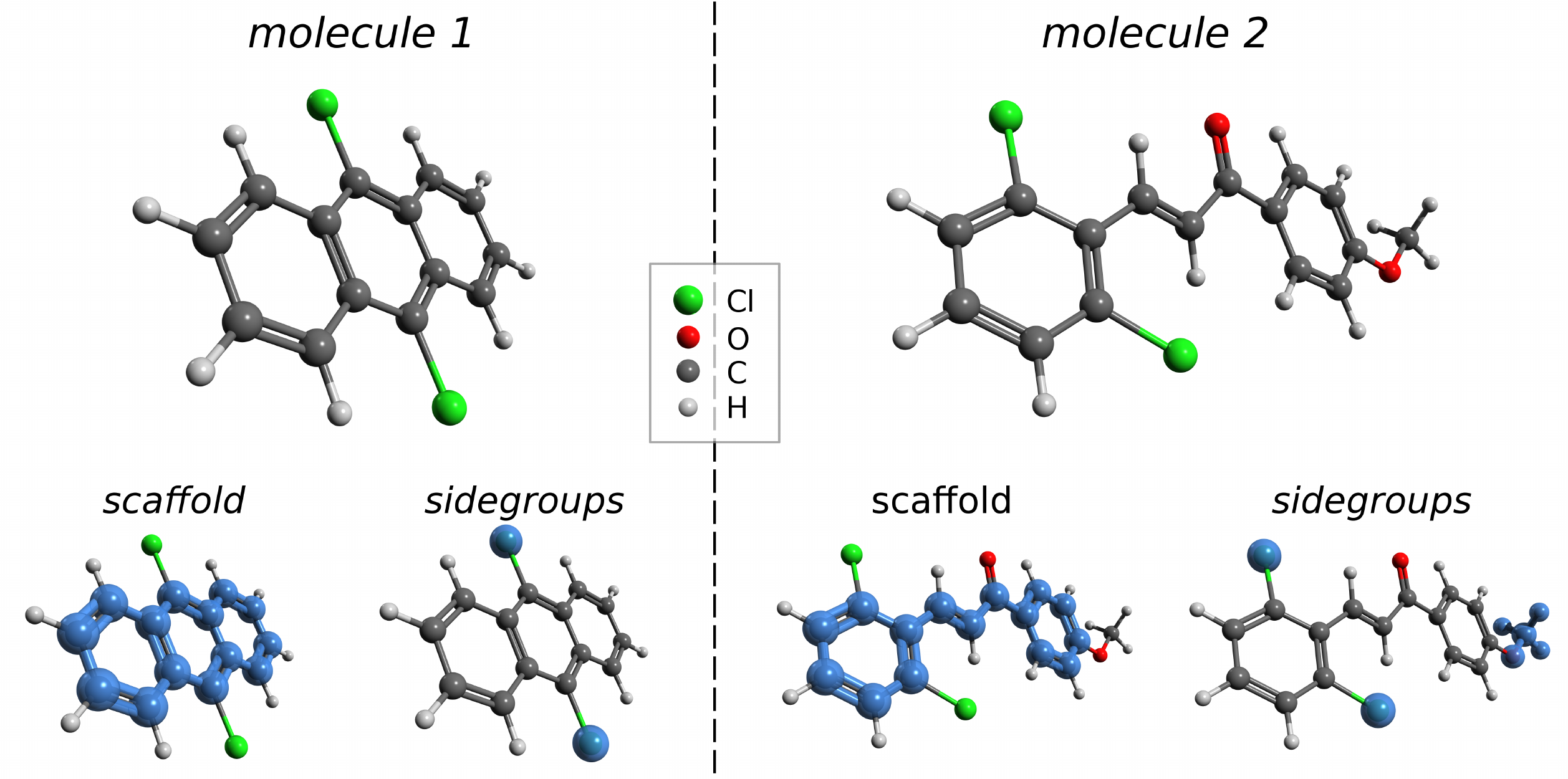}
  \caption{Illustration of the Bemis-Murcko partitioning scheme. For two example molecules, the decomposition into scaffold 
	and sidegroups is illustrated by highlighting the scaffold in light blue on the left and by highlighting the sidegroups on 
	the right.}
  \label{fig:BM-cluster_example}
\end{figure}

Compared to earlier, focused studies on OSCs, the dataset used in this study covers a wide range of molecular diversity to ensure the generality of our data mining results. This is reflected by 83 crystal space groups occuring at least 3 times in the dataset, while the mono-molecular building blocks are correspondingly distributed over more than 20 molecular point groups. The molecules comprise H, B, C, N, O, F, Si, P, S, Cl, Se, Te, Br, I, and As species and therewith cover the entire range of elements commonly found in OSCs. As further specified below, a variety of molecular backbones is contained in the database, featuring single to multiple (hetero)cycles. These backbones are functionalized with 4229 different sidegroups, 828 of which occur in more than three different 
crystals. 

In order to extract correlations from such a diverse set of data, a suitable representation is required that collapses its dimensionality to a tractable size. In contrast to more abstract substructure or statistical learning based approaches \cite{scaffold-network}, we here pursue a chemically intuitive concept that partitions the molecular building block of the organic crystal into ring systems, linkers and sidegroup atoms. The Bemis-Murcko (BM) scaffold \cite{BM-Paper} is thereby specifically defined as the molecular core comprised of connected ring systems and their linkers, resulting after removal of all sidegroup atoms. Figure~\ref{fig:BM-cluster_example} provides two illustrative examples of this decomposition scheme. While BM scaffolds have so far mainly been used in medicinal chemistry to study properties of drug-like molecules\cite{scaffold-review1, scaffold-review2, BM-example2}, the concept is also well-suited for our purpose. It mainly extracts $\pi$-conjugated ring systems which to date almost exclusively form the basis of molecular electronics.\cite{pi-systems, heteroarenes} The scaffold also constitutes the largest part of the molecule which is likely most decisive in terms of electronic structure, shape, conformational flexibility\cite{compound-set-enrichment} and possibly geometric arrangement in the crystal. Consequently, we disregard some structures in the ``64k-dataset'' where very extensive sidegroups constitute more than 50 \% of the non-hydrogen atoms. 

Using this BM representation allows us to cluster the dataset entries based on their respective scaffolds. In order to obtain representative sample sizes, we required each such cluster to contain at least $15$ crystals. Although this rigorous filtering discards many clusters due to too small cluster sizes, it leads to a well-analyzable subset of the ``64k-dataset'' that contains $7.569$ molecular crystals distributed over $195$ scaffold clusters. These molecular crystals are made of $6.936$ unique molecules, since $633$ crystals appear as polymorphic structures or had entered the Cambridge Structural Database more than once, e.g.~coming from different reported experimental sources. Nevertheless, each scaffold cluster comprises at least $10$ unique molecules and we provide a list of BM cluster scaffolds, the detailed distributions of cluster sizes, as well as $|H_{ab}|$ and $\lambda$ values in this subset in the Supporting Information (SI) Figs. S2-4. For the remainder of the study we will focus on this data subset and will consistently refer to it as our database.

\begin{figure}
    \includegraphics[width=.6\textwidth]{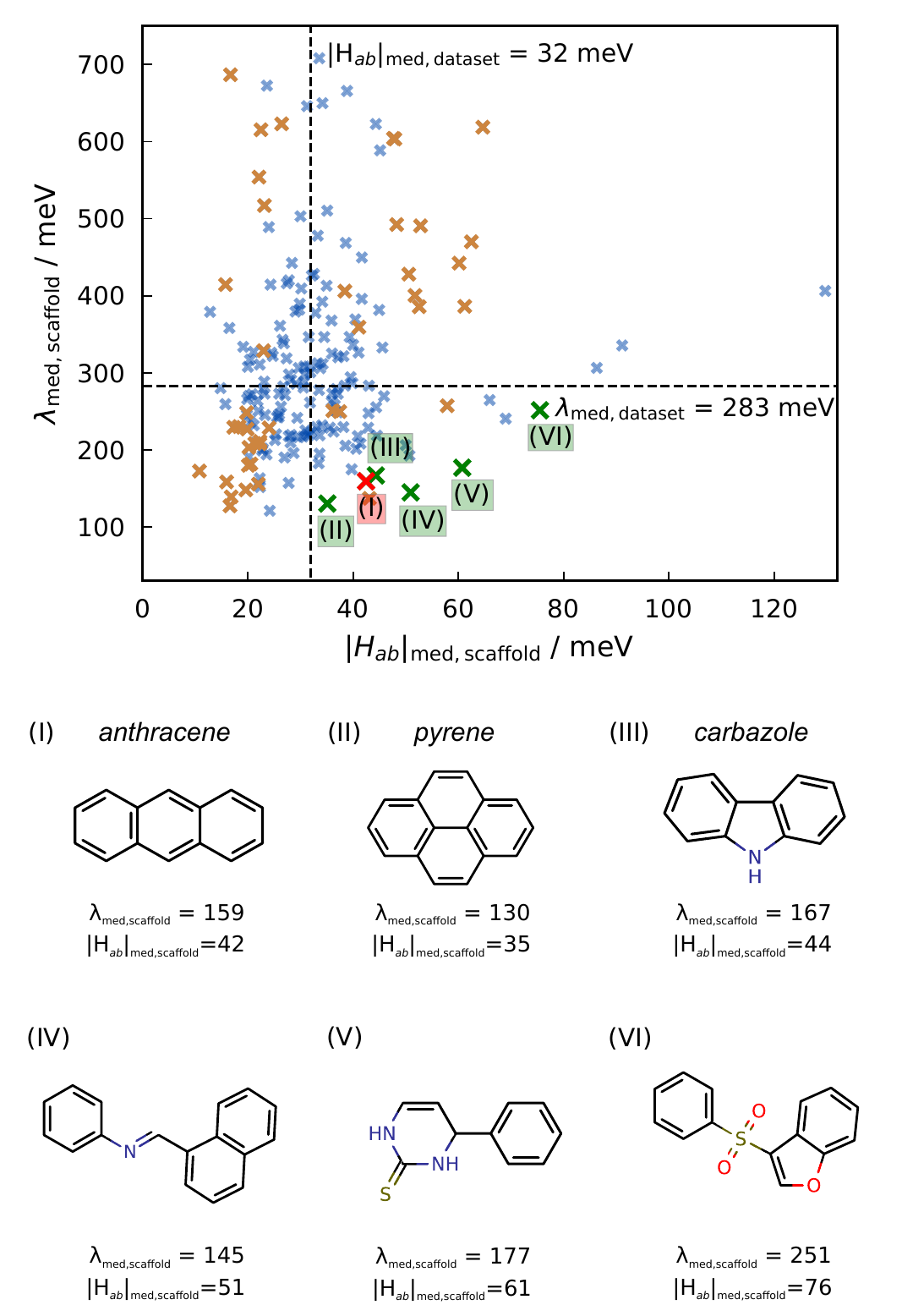}% \\
  \caption{Top: Scatterplot displaying the median $|H_{ab}|_\text{med,scaffold}$ and $\lambda_\text{med,scaffold}$
	for each of the $195$ scaffold clusters. Orange crosses mark clusters with a $|H_{ab}|$ and $\lambda$ distribution 
	significantly differing from each of the respective background distributions, and blue crosses mark clusters with no such significant difference. The medians of the two descriptors over the entire
	dataset are marked with dashed lines. The anthracene cluster is highlighted in red as a reference cluster containing well-known 
	high charge-carrier mobility crystals. Bottom: Examples of BM-scaffolds further discussed in the text. 
	The datapoints for these scaffolds are labeled and depicted in green color in the top panel.}
  \label{fig:BM-Scatterplot-scaffold-examples}
\end{figure}

In Fig.~\ref{fig:BM-Scatterplot-scaffold-examples} we first analyze the median over $|H_{ab}|$ and $\lambda$ for each scaffold cluster and contrast it with the corresponding medians over the entire database. This provides a chemically intuitive assessment of the performance and tuning potential of each particular scaffold. Within the scope of structures in each scaffold cluster, this averaging step accounts for changes in electronic and crystal structure, brought about by differing anchor points of the varying sidegroups, thus to a certain extent accounting for the so called positional disorder\cite{LehnherrOrgLett2012} in each scaffold.

To provide a reference of a well-performing scaffold, we highlight in Fig.~\ref{fig:BM-Scatterplot-scaffold-examples} the anthracene scaffold cluster. With anthracene a well-known base material for organic electronics\cite{anthracenes}, this reference cluster is predominantly composed of high charge-carrier mobility crystals with relatively low $\lambda$ and high $|H_{ab}|$. Compared to this reference cluster, there is a wide spread of the median descriptors among the 195 scaffold clusters. In particular, there are many scaffolds with exceedingly poor median charge-transport characteristics (very high $\lambda_\text{med,scaffold}$ and/or very low $|H_{ab}|_\text{med,scaffold}$). On the other hand, there are also several scaffold types that exhibit at least as favorable characteristics as the anthracene reference cluster. Note, that in particular improvements in the reorganization energy, i.e.~a decrease, can thereby also readily compensate for a deterioration in the electronic coupling (or vice versa). Within a Marcus model, a $50\%$ improvement (deterioration) of $\lambda$ could for instance tolerate a decrease (increase) of $|H_{ab}|$ by up to $18$ meV ($26$ meV) to still yield the same 
hopping rates, and by implication mobilities, as anthracene. Essentialy all of the scaffolds highlighted in Fig.~\ref{fig:BM-Scatterplot-scaffold-examples} fall in approximately the same range of mobility. Considering the immense range of $\lambda_\text{med,scaffold}$ values included in our dataset and comparing it to the narrow range of available $|H_{ab}|_\text{med,scaffold}$ values, we again find that the reorganization energy is an ideal first level target for molecular optimization.
The large spread of $\lambda_\text{med,scaffold}$ seen in Fig.~\ref{fig:BM-Scatterplot-scaffold-examples} over all scaffolds indicates thereby the vastly different degrees of inner-sphere reorganization caused by the different molecular cores. A spread along $|H_{ab}|_\text{med,scaffold}$, on the other hand, can be caused by a number of factors, such as varying capabilities of scaffolds to stabilize their most favorable packing motives, as well as intrinsic differences of the involved frontier orbital geometries.\cite{Oberhofer2017CR}

In order to assess the statistical significance of this observed spread in the median descriptors we employ non-parametric statistical testing based on the two-tailed Mann-Whitney-U test.\cite{Mann1947} As further detailed in the methods section, this test determines non-parametrically the probability $p$ of two sets of values to belong to the same distribution. By correcting detection bias due to the relatively large number of tests being carried out, this yields the $q$ probability and we judge from a value $q < 5$\,\% that the distribution of descriptor values in a given scaffold cluster differs significantly from the distribution in the whole database. With respect to the $\lambda$ descriptor, $105$ clusters, i.e. more than half of all considered scaffolds, fulfill this restrictive criterion. With respect to the $|H_{ab}|$ descriptor, this number is a bit lower, but with $69$ still extends over more than a third of the 195 scaffolds (the full data for all scaffolds is given in Table S1 in the SI). A total of 44 clusters exhibits statistically different distributions in both descriptors and the corresponding data is color-coded in Fig.~\ref{fig:BM-Scatterplot-scaffold-examples}. The statistical analysis therewith confirms our initial notion that BM scaffolds are a suitable data representation that allows to differentiate molecular crystals in terms of their charge-transfer properties. Favorable scaffolds with a general tendency to high electronic couplings and low reorganization energies can thus be identified from the available data. Smaller such scaffolds featuring one or two aromatic rings could then be used as promising building blocks, while larger scaffolds can serve as a starting point for further functionalization. In Fig.~\ref{fig:BM-Scatterplot-scaffold-examples} we specifically highlight a few such identified smaller scaffolds that we further discuss in Section III below.

\subsection{Molecular Sidegroups}

Having identified suitable molecular scaffolds, we next turn our attention to the sidegroups. The introduction of appropriate sidegroups has been shown to improve carrier mobilities by influencing the stable crystal structure\cite{Liu2018ACR, hydrogen-bonded-interactions, Sosorev2017PCCP} or by suppressing vibrationally induced disorder\cite{reducing_dynamicdisorder, TCNQ_dynamic_disorder}. Here, we focus specifically on the influence on the reorganization energy. Previous work has confirmed this critical influence in comparative studies attaching different sidegroups to a given single scaffold.\cite{nonbonding_lambda_1, substitution_effects_bredas, oshi-naphthalene-substitution, oshi-anthracene-tetracene-substitution, Lee2010, Hutchison2005JACS, VEDOVA-BROOK2004-Chem_Phys, ZHAO201325} Building on this empirical concept, we now use our extensive database to investigate whether there are molecular sidegroups that generally lower the reorganization energy, independent of the scaffold they are attached to. We achieve this with the same techniques of statistical analysis as before for the scaffolds, i.e. we first identify all sidegroups in the database and then cluster the data. This time, one cluster corresponds to all molecules (of any scaffold type) that contain a given sidegroup. This does not differentiate between specific attachment positions to the scaffold or the coexistence of multiple sidegroups in one molecule, and we return to this point below. To ensure meaningful statistics, we only consider sidegroup clusters comprising at least 10 different molecules, with the notable exception of the \ce{-Se-CH3} sidegroup. This sidegroup occurs only in six molecular crystals in the database, but is nevertheless considered because of its exceptional properties ({\em vide infra}). Altogether this yields $70$ sidegroup clusters containing a total of $6.162$ unique molecules.
\pagebreak
\begin{figure}[ht]
 \centering
   \vspace{-7px}
  \includegraphics[width=.65\textwidth]{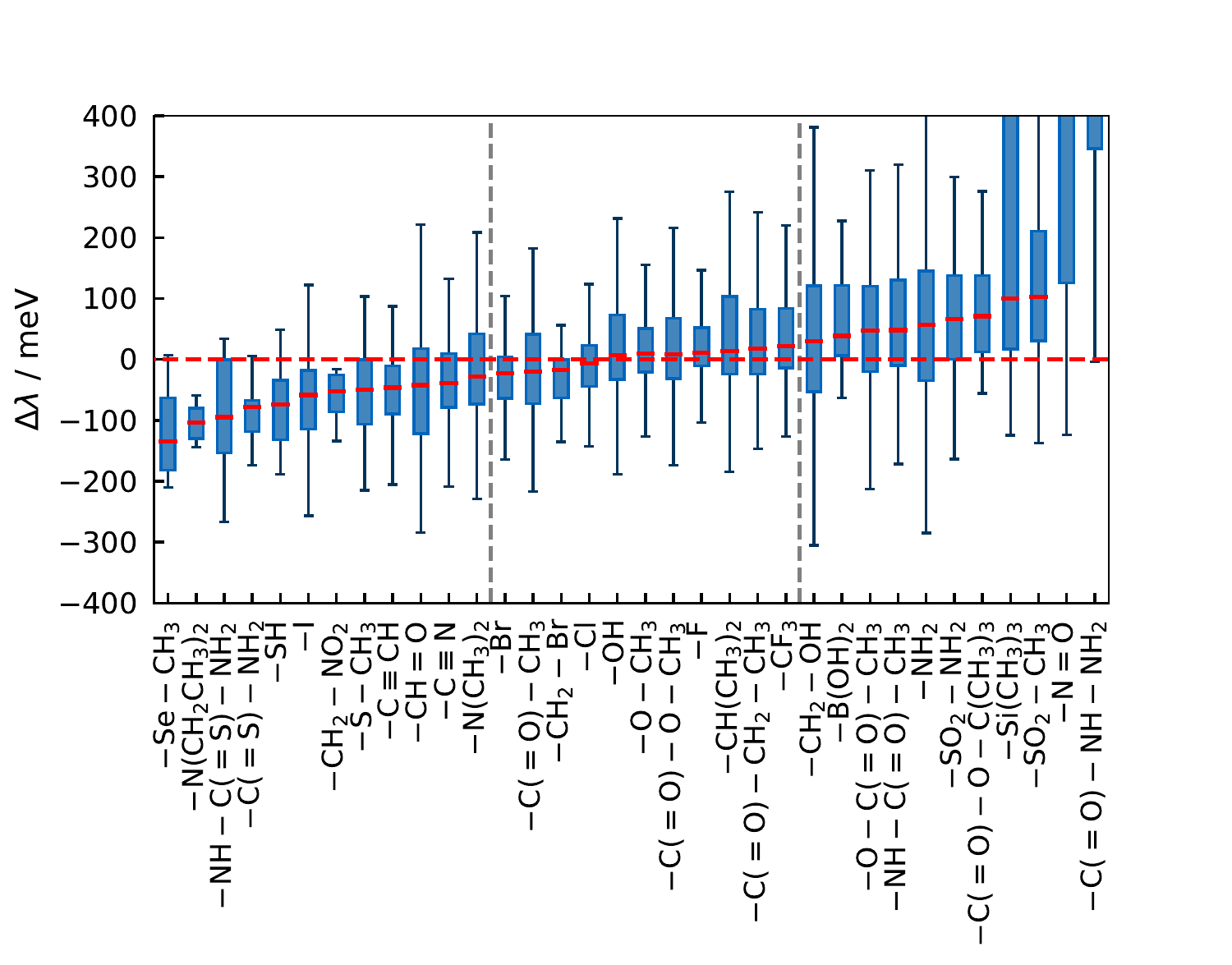}
  \\
  \vspace{9px}
  \includegraphics[width=.63\textwidth]{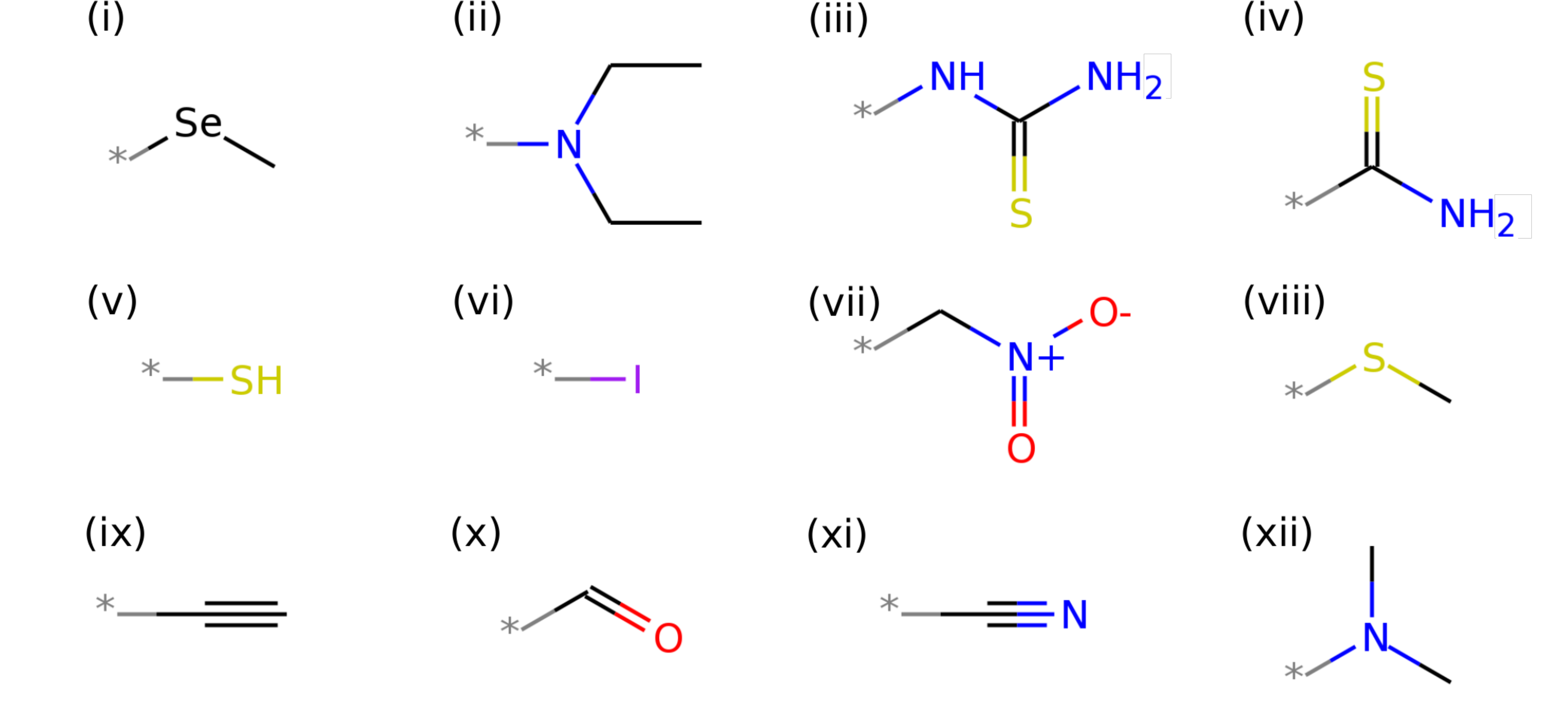}
  \caption{Top: Boxplot showing the distributions of relative reorganization energies $\Delta\lambda$ of $34$ sidegroups,
	see text. They are sorted by their median $\Delta \lambda_\text{med,sidegroup}$ which is marked by a solid red line. Blue 
	boxes extend from lower to upper quartiles of the data, while the vertical lines (whiskers) delineate the most extreme points.
  Gray-dashed vertical lines serve as separators to distinguish sidegroups that generally improve (i.e. lower)
	the reorganization energy ($\Delta \lambda_\text{med,sidegroup} < -25$,meV, to the left in the plot) and sidegroups that 
	generally worsen (i.e. increase) the reorganization energy ($\Delta \lambda_\text{med,sidegroup} > 25$,meV, 
	to the right in the plot). Due to space constraints, the labels for the different sidegroups use (=O/S) to denote branches, 
	similar as is done in canonical smiles strings \cite{Weininger1989}. Bottom: Identified 12 most favorable sidegroups 
	appearing on the left hand side of the boxplot.}
	\vspace{-5px}
  \label{fig:boxplot_sidegroup_evaluation}
\end{figure}
\pagebreak

For each molecule in a given sidegroup cluster we determine its relative reorganization energy $\Delta\lambda$ as the difference between the reorganization energy $\lambda$ of the molecule and the median reorganization energy $\lambda_\text{med,scaffold}$ over the cluster of molecules with this particular scaffold, i.e. the $\lambda_\text{med,scaffold}$ evaluated in the last section. Thus, a negative value of $\Delta\lambda$ indicates that the attachment of the sidegroup to the scaffold leads to compounds of lower reorganization energy as compared to the attachment of other sidegroups to the same scaffold. Note that due to the diversity of the database it is possible that certain sidegroups appear multiple times on a single scaffold. In order not to over-emphasize such cases, we divide each $\Delta\lambda$ by the number of occurrences. Taking the median of all $\Delta\lambda$ within one sidegroup cluster, we finally arrive at $\Delta\lambda_\text{med,sidegroup}$, which evaluates how the attachment of the sidegroup generally influences the reorganization energy over different scaffold types. 

For many sidegroups, the obtained $\Delta\lambda_\text{med,sidegroup}$ differ significantly from zero. This indicates that these sidegroups indeed either consistently worsen or consistently improve the reorganization energy, regardless of the scaffold they are attached to. Mann-Whitney-U testing \cite{Mann1947} is again employed to verify that these results are not due to chance or a biased sample selection. For $34$ of the $70$ sidegroup clusters this statistical testing confirms that their relative reorganization energy distributions differ significantly from the distribution in the entire database ($q < 5$\,\%). In particular for the \ce{-Se-CH3} functional group a very low $q$-value of 1\,\% confirms the statistical relevance despite the smaller sample size.

Figure \ref{fig:boxplot_sidegroup_evaluation} compiles the $\Delta\lambda$ distributions of these $34$ sidegroup clusters. Groups appearing at the left side of Fig.~\ref{fig:boxplot_sidegroup_evaluation} tend to reliably decrease $\lambda$, while those on the right tend to increase the reorganization energy. As apparent, a $\Delta\lambda$ distribution within a sidegroup cluster differing significantly from the distribution over the entire database does not necessarily mean that the median $\Delta\lambda_\text{med,sidegroup}$ has to differ markedly from zero. For the corresponding sidegroups in the middle of Fig.~\ref{fig:boxplot_sidegroup_evaluation}, the effect of the sidegroup seems more subtle, sometimes increasing, sometimes decreasing the reorganization energy. We speculate that in these cases the (presently unresolved) specific attachment position or the coexistence of other sidegroups could play a decisive role \cite{substitution_effects_bredas}. More importantly, there are, however, $12$ sidegroups which on average decrease the reorganization energy by more than $25$\,meV, see Fig. \ref{fig:boxplot_sidegroup_evaluation}. These sidegroups are obviously good candidates for $\lambda$-tuning approaches and will be further discussed in Section III below.

\subsection{Sidegroup Engineering}

Our data analysis to this point has identified suitable scaffolds that yield generally high $|H_{ab}|$ and low $\lambda$ values, as well as suitable sidegroups that consistently lower $\lambda$. This suggests an appealing design principle to arrive at new molecular crystals with optimized charge-transfer properties: Select a favorable scaffold with high $|H_{ab}|$ and further optimize the reorganization energy by addition of favorable sidegroups or by exchange of unfavorable with favorable sidegroups. Unfortunately, this neglects the known sensitive influence of sidegroups on the crystal structure.\cite{mmpa_csd, Alberga2017CM} In contrast to $\lambda$, which is predominantly determined by the intramolecular structure, $|H_{ab}|$ describes the electronic coupling between molecules in the crystal and is thus critically dependent on the crystal structure. While sidegroup tuning could improve the reorganization energy, it could also change the crystal structure and then potentially worsen $|H_{ab}|$.

To address this aspect we use matched molecular pair (MMP) analysis \cite{mmpa_in_short, MMPA} to identify those molecules in the database, where the resulting molecule after addition or exchange of a single sidegroup is still contained in the database. This way, the crystal structure of both the initial and the modified molecule is available, enabling us to explicitly analyze the influence of the sidegroup engineering on the electronic coupling $|H_{ab}|$. Specifically, we thereby focus on the exchange of an unfavorable sidegroup ($\Delta\lambda_\text{med,sidegroup} > 0$\,meV) with a sidegroup identified as very favorable above ($\Delta \lambda_\text{med,sidegroup} < -25$\,meV), cf. Fig. \ref{fig:boxplot_sidegroup_evaluation}. This generates a testset S1 containing 206 optimized molecules. As a control series, we also generated a testset S2 containing $124$ molecules, where a favorable sidegroup ($\Delta\lambda_\text{med,sidegroup} < 0$\,meV) is replaced with a highly unfavorable one ($\Delta \lambda_\text{med,sidegroup} > 25$\,meV). In analogy, testset S3 with $161$ molecules (S4 with $91$ molecules) contains molecules where a highly favorable (unfavorable) sidegroup has been added to the molecular scaffold. The size discrepancy between the four testsets is due to the different availability of experimental crystal structures. A full description of the testsets and the ensuing analysis is provided in the SI.

\begin{figure}
  \includegraphics[width=.75\textwidth]{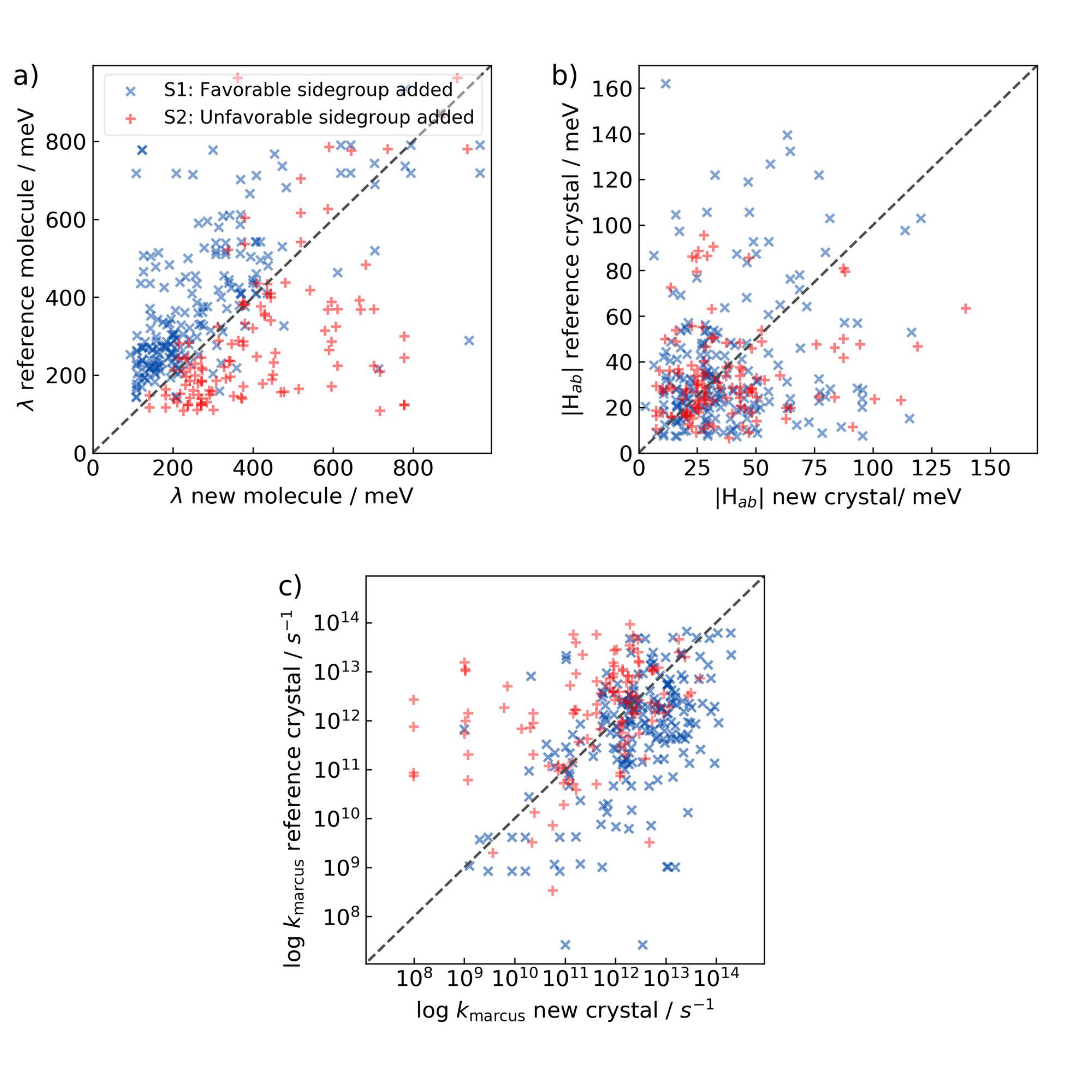}
  \caption{Charge-transfer parameter tuning by sidegroup exchange for the series of molecules contained in the ``optimized'' testset
	S1 (blue data points) and the control testset S2 (red data points), which was deliberately made worse. Shown are changes in 
	a) $\lambda$, b) $|H_{ab}|$ and c) Marcus type hopping rate $k_{\rm Marcus}$ 
	when going from a reference molecule to a new molecule (see text).}
  \label{fig:sidegroup-MMPA1}
\end{figure}

Fully supporting our sidegroup analysis above, we obtain marked improvements of the reorganization energy (up to changes of a factor of 5) for the overwhelming majority of modified crystals in the two ``optimized'' testsets (87\% of all crystals in S1, 77\% in S3), see Fig. \ref{fig:sidegroup-MMPA1}. Similarly, the ``worsened'' control sets S2 and S4 show predominantly a large deterioration of the $\lambda$ (in 80\% of cases in S2, 73\% in S4). As expected, the sidegroup engineering also influences the electronic coupling. However, the changes there are mostly unsystematic. In the ``optimized'' testsets S1 and S3  we find a roughly equal number of cases where $|H_{ab}|$ is improved to those where the coupling deteriorates. Interestingly, in the ``worsened'' testsets S2 and S4, we still obtain a slight improvement of $|H_{ab}|$ in 58 \% and 59 \% of all cases, respectively. This is most likely attributed to selection bias, as the same correlation of large couplings with regrettably large reorganization energies is already present in the molecules of ``64k-dataset''\cite{Schober2016JPCL}. Notwithstanding, in particular for the relevant larger coupling values, the induced unsystematic changes are comparatively small. As the reorganization energy furthermore enters e.g.~the Marcus model exponentially, this suggests that these changes in $|H_{ab}|$ should not break the overall trend. Indeed, when we compute the Marcus transfer rates $k_{\rm Marcus}$ on the basis of the obtained couplings and reorganization energies,~\cite{Oberhofer2012PCCP} we still find increased rates in 71\% (61\%) of the molecular crystals in the ``optimized'' testset S1 (S3), and decreased rates in 67\% (59\%) of the molecular crystals in the ``worsened'' control testsets S2 (S4). This suggests that sidegroup engineering is (at least to a certain extent) robust with respect to induced changes in the crystal structure and is therefore a viable approach to optimize the charge-transport properties of molecular crystals.

\section{Discussion}

In this work we explored the possibility of optimizing organic crystals for charge transport using a data-driven approach. Mining a large database of two major charge-transport descriptors, the reorganization energy $\lambda$ and the electronic coupling $|H_{ab}|$, for close to 7.000 mono-molecular crystals, we transferred the concept of Bemis-Murcko scaffolds from pharmaceutical research to cluster the data according to molecular scaffolds and the attached sidegroups. For both types of clusters we obtain statistically confirmed structure-property relationships, i.e. we identify scaffolds or sidegroups that are generally favorable for charge transport.

A number of promising scaffolds with generally low $\lambda$ and high $|H_{ab}|$ regardless of their functionalization are identified in the lower right quadrant of Fig.~\ref{fig:BM-Scatterplot-scaffold-examples}. Among them are scaffolds contained in well-known OSCs: anthracene\cite{anthracenes} (Fig.~\ref{fig:BM-Scatterplot-scaffold-examples} I), pyrene\cite{pyrenes} (Fig.~\ref{fig:BM-Scatterplot-scaffold-examples} II), or carbazole\cite{Reig2017PCCP} (Fig.~\ref{fig:BM-Scatterplot-scaffold-examples} III). Based on our strict statistical significance criterion, all three exhibit consistently improved charge-transport parameters compared to the database background. The only exception is $|H_{ab}|_\text{med,scaffold}$ of the pyrene scaffold II, which shows only little statistical significance due to the limited occurrence of this scaffold in our database.

Next to these more common aromatic systems, there are also a few promising linked ring systems like scaffolds IV and V shown in Fig.~\ref{fig:BM-Scatterplot-scaffold-examples}, which one would not immediately associate with the OSC context. They do, however, exhibit much higher median electronic couplings and comparable reorganization energies as the anthracene reference cluster. Multiple other scaffolds are particularly appealing with regard to only one of the two charge-transport descriptors. In particular, scaffolds like Fig.~\ref{fig:BM-Scatterplot-scaffold-examples} VI exhibit extraordinarily high electronic couplings suitable for organic electronics, but feature on average unsuitably high reorganization energies. These scaffolds are naturally of specific interest for a targeted optimization through functionalization with favorable sidegroups.

Our data-driven analysis of sidegroups paired with tests of statistical significance allows for a broader perspective on existing empirical strategies of sidegroup tuning. Among the $34$ identified sidegroups that significantly alter the reorganization energies of their respective scaffolds are sidegroups like \ce{-OH}, \ce{-O-CH3}, \ce{-NH2}, \ce{-F}, \ce{-Cl}, \ce{-CF3}, or \ce{-C#N} that had been considered in preceding comparative studies \cite{nonbonding_lambda_1, substitution_effects_bredas, oshi-naphthalene-substitution, oshi-anthracene-tetracene-substitution, Lee2010, Hutchison2005JACS, VEDOVA-BROOK2004-Chem_Phys, ZHAO201325, PAN2006CompMaterSci, JANPRAPA2018ChemPhys}. In parts, these earlier studies found trends that match the general structure-property relationships established in this work, for example for \ce{-C#N}, \ce{-F}, \ce{-CF3}, and \ce{-NH2} substitutions on thiophenes, furans, and pyrroles.\cite{Hutchison2005JACS}. Interestingly, with the exception of the cyano group, these often studied sidegroups not always seem to have a favorable effect on reorganization energies, with some (\ce{-CF3}, and \ce{-NH2}) even showing---on average---the opposite trend. The here confirmed favorable influence of the cyano group has been explained in the literature\cite{nonbonding_lambda_1} by the introduction of a local nonbonding character in the frontier orbitals. In contrast, the common and much studied class of alkyl sidegroups is, with the exception of \ce{-CH(CH3)2}, not represented among the $34$ sidegroups with significant influence. On the basis of our present database we can thus not confirm any systematic improvement or deterioration of the reorganization induced by this class of sidegroups. Instead and most intriguingly, many of the here identified $12$ sidegroups that lead to a marked improvement of $\lambda$ have to the best of our knowledge only rarely been discussed in the OSC context. This holds for \ce{-Se-CH3}, \ce{-N(CH2-CH3)2}, \ce{-SH},\cite{ZHAO201325} \ce{-I}, \ce{-CH2-NO2}, \ce{-S-CH_3}\cite{JANPRAPA2018ChemPhys}, \ce{-C#CH},\cite{ZHAO201325}, \ce{-CH=O},\cite{ZHAO201325} or \ce{-N(CH3)2} \cite{PAN2006CompMaterSci, JANPRAPA2018ChemPhys}, where we added the few references to works addressing these sidegroups we could find at all. A detailed analysis of the reason for the consistent $\lambda$ reduction induced by these sidegroups is presented in the SI. It confirms that each of these sidegroups yields first of all an expected additional contribution to the reorganization energy.\cite{substitution_effects_bredas} However, this contribution is more than compensated by a reduction of the scaffold reorganization energy. Especially the best performing \ce{-Se-CH3} sidegroup achieves this by efficiently delocalizing the charge throughout the scaffold and the sidegroup.

The clear structure-property relationships for both scaffolds and sidegroups suggest the possibility of a targeted molecular design by combining favorable scaffolds with reorganization energy-lowering sidegroups. This effect is expected to lead to improved charge transport rates, with the reorganization energy entering e.g.~a Marcus-like\cite{Marcus1993RMP} model exponentially, while $|H_{ab}|$ only enters quadratically.\cite{Oberhofer2012PCCP} Our data analysis indicates that such a design strategy is robust (at least to a certain extent) to a change in the crystal structure that might be induced by such functionalization. Accordingly, we would presently predict the addition for instance of most favorable \ce{-Se-CH3} or \ce{-N(CH2-CH3)2} sidegroups to any of the favorable scaffolds highlighted in Fig.~\ref{fig:BM-Scatterplot-scaffold-examples} to yield highly promising OSC candidates for experimental synthesis. Much of the uncertainty in these predictions could be removed by explicitly validating that the electronic couplings remain indeed favorable in the crystal structure adopted by the corresponding compound. Unfortunately, computational crystal structure prediction is still a highly challenging task.\cite{Maddox1988NAT,Reilly2016ACSB} In fact, the limitations in rapidly and reliably determining the crystal structure for a larger number of promising candidates represent in our view currently the major bottleneck to a large-scale data-driven {\em in silico} discovery of improved OSCs. This prevents a targeted extension of the database through additional first-principles calculations, and correspondingly restricts the data mining to compounds for which experimental crystal structures are available in the Cambridge Structural Database \cite{CSD}. The concomitant relative scarcity of data also prohibits refined data clustering that notably would distinguish different anchor points of sidegroups to a scaffold or the coexistence of multiple sidegroups. Efficient crystal structure prediction and the concerted buildup of experimental and computational structural databases are therefore key to unleash the full potential of modern data mining or machine learning approaches for organic semiconductors, and thereby navigate the vast and largely unexplored design spaces of this intriguing class of molecular materials.

% \\
\begin{suppinfo}
Additional detailed information on the BM scaffold dataset: Distributions of $|H_{ab}|$, $\lambda$ over the dataset, as well as a distribution of the number of molecules in the scaffolds and a list of all scaffolds. In addition, an extended boxplot for all sidegroup clusters as well as a depiction of all identified sidegroups is provided. A series of \ce{-H}, \ce{-S-CH3} and \ce{-Se-CH3} exchanges on different molecules further illustrates the favorable impact of the latter two groups on gas-phase reorganization energies, while a detailed discussion of the underlying factors is provided for one example.
\end{suppinfo}

\begin{acknowledgement}
We want to thank Lynne Stecher for very helpful discussions on the statistical methods. We further acknowledge support from the Solar Technologies Go Hybrid initiative of the State of Bavaria and the Leibniz Supercomputing Centre for high-performance computing time at the SuperMUC facility.
\end{acknowledgement}

\bibliography{main}

\end{document}